# Hybrid Modeling Application in Control Valve

Yuan Chi[1], He Xu[1] *, Feng Sun[1] and Yufeng Qian[2]

[1] College of mechanical and electrical engineering Harbin Engineering University , 150001, China
[2] Jiangsu Yi-valve Co. LTD Yang Zhong ,212200, China

**ABSTRACT**: In view of the serious nonlinearity, time-varying and parameter uncertainty in the physical model of regulating valve, a prediction model of flow rate and pressure of regulating valve based on mixed model was proposed.According to the physical model of the regulator, the parameters that can represent the operation state of the regulator are analyzed, and the relevant parameters are identified by unbiased LSSVM method.The DAMADICS simulation results show that the model can predict the output of flow rate and pressure with better accuracy, which can provide guidance for the design of automatic control valve or fault diagnosis system.

## 1 INTRODUCTION

The regulator is the most important terminal element in the process control industry. The establishment of an accurate mathematical model is the key to the model-based automatic control or fault diagnosis system. The common modeling methods of the regulator include mechanism analysis method, simulation analysis method and data-driven system identification.Due to the complex structure, strong nonlinearity, hysteresis, dead zone and other factors, it is difficult to establish the analytical model of the control valve by using the laws of physics, so the simulation analysis method and the system identification method based on data drive are more applied.Based on data driven approach in modeling process using the data in the actual work process, do not need to consider the internal structure and the laws of physics, as the detection technology, the rapid development of computer technology, artificial intelligence technology, regulating valve through system identification modeling is more easy to implement, higher reliability, so this method be regulator research hotspot in the field of modeling.Neural network algorithm is the most used, but with the development of research, it is found that neural network has a large required data sample, easy to appear low learning or generalization ability control valve fluid dynamics model[1,2].

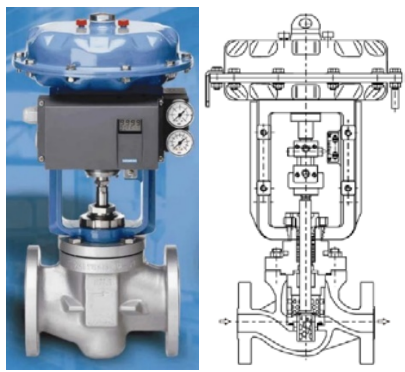

Fig. 1 Model of pneumatic thin film single-seat through regulating valve

Regulator consists of actuator, body and positioner.According to the types of energy used by the actuator, it can be divided into electric actuator, hydraulic actuator and pneumatic actuator.This paper analyzes the pneumatic straight single-seat control valve, as shown in Figure 1, which is the control valve studied in this paper[3].

## 2 FLUID DYNAMICS MODEL OF CONTROL VALVE

The simplified structure of the regulating valve body is shown in Figure 2. By controlling the flow area of the control medium moving up and down the valve core, the flow and pressure of the fluid and other physical quantities can be controlled. The regulating effect of the regulating valve can be analyzed according to the thin-wall orifice theory.[4]

$$Q = C_v \varepsilon \frac{A_c}{\sqrt{1-\beta^4}} \sqrt{\frac{2(p_1 - p_c)}{\rho_1}} \quad (1)$$

Where, $Q$ is the volume flow rate; $C_v$ is the flow velocity coefficient; $\varepsilon$ is the expansibility coefficient; $A_c$ is the flow area of the section at the shrinkage; $p_1$ and $p_c$, respectively, are pressure in front of the valve and at the shrinkage (C-C section); $\beta$ is the diameter ratio (equivalent diameter $d_e$ of the section at the shrinkage, and the ratio of equivalent diameter $d_1$ of the section of the pipeline); $\rho_1$ is the average density of fluid at section 1-1 in front of the valve.

The flow coefficient represents the flow capacity of the regulator, which is related to the structure of the valve core and seat, the pressure difference between the regulator and the valve seat, and the nature of the fluid. Regulating valve is essentially to control the signal to change the stroke of the valve stem, so as to change the flow area, change the resistance coefficient and achieve the purpose of regulating the flow. Structure of benchmark actuator system is shown in Fig 2[5].

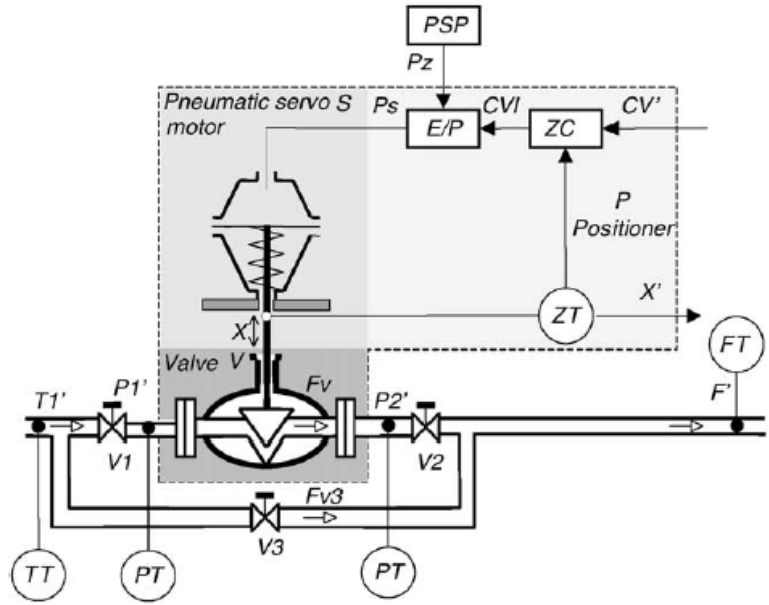

Fig. 2. Structure of benchmark actuator system.

At the vena contracta (smallest cross sectional area of the flow stream) the fluid pressure is minimal while the velocity is maximal. If the pressure at the vena contracta falls below the vapour pressure of the liquid, vapour cavities forms in the flow stream. Cavitation results if fluid outlet pressure p2 recovers to a pressure above the vapour pressure of the liquid. As the vapour cavities collapse, noise is generated and damage can occur. Cavitation damage produces a rough, pitted, cinder-like surface. If the fluid outlet pressure p2 remains below the vapour pressure of the liquid, flashing effect occurs. Flashing damage resembles erosion and is distinguished by the smooth polished appearance of the eroded surface.

Pressure recovery coefficient

Pressure recovery coefficient $F_L$ is defined as follows

$$F_L = \sqrt{\frac{p_1' - p_2'}{p_1 - p_{vc}}} = \sqrt{\frac{\Delta p_T}{p_1 - F_F P_v}} \quad (2)$$

Where, $p_1' - p_2'$ and $\Delta p_T$ are the pressure difference at both ends of the regulating valve, unit KPa: $p_{vc}$ is the pressure at the point where shrinkage occurs, unit KPa, $F_F$ is the critical pressure ratio; $P_V$ is liquid saturated vapor pressure (KPa);

Liquid saturation vapor pressure

The relationship between the saturated vapor pressure and temperature of the liquid can be expressed by clausius-Clapeyron equation

$$\frac{d \ln p_v}{dT} = \frac{\Delta_{vap} H_m}{RT^2} \quad (3)$$

Where, R is the molar gas constant, and its value is 287 $N \cdot m / (kg \cdot K)$; T is the thermal temperature, unit K, and $\Delta_{vap} H_m$ is the molar heat of vaporization of pure liquid at the temperature plant, unit J/kg.

And since $\Delta_{vap} H_m$ is going to change less for a smaller temperature range, we can approximate $\Delta_{vap} H_m$ as a constant. So you integrate (3), you get

$$In p_v = -\frac{\Delta_{vap} H_m}{R} \cdot \frac{1}{T} + C \quad (4)$$

In the formula, C, R and $\Delta_{vap} H_m$ are all constants. So, $In p_v$ is a line R with a slope of minus $-\frac{\Delta_{vap} H_m}{R}$.

critical pressure ratio $F_F$

Different media have different degrees of deviation from $P_V$ when blocking flow is generated. According to the experimental results, the critical pressure ratio $F_F$ is related to the ratio of saturated vapor pressure $p_v$ and critical pressure. The value of $F_F$ can be approximately calculated as follows:

$$F_F = 0.96 - 0.28\sqrt{\frac{p_v}{p_c}} \qquad (5)$$

Where, $p_c$ is the critical pressure, unit KPa; For water $p_c = 22565 KPa$

According to the calculation method of flow coefficient of each fluid state, it is summarized as follows

(a) Non-blocking turbulence. The formula is derived from the basic formula in GB/T 17213.1-2015;

Application conditions: $\Delta p < F_L^2 (p_1 - F_F p_v)$

The flow coefficient shall be determined by the following formula

$$C = \frac{q_v}{N_1} \cdot \sqrt{\frac{\rho_1/\rho_0}{\Delta p}} \qquad (6)$$

Note: The numerical constant $N_1$ depends on the type of unit and flow coefficient $K_v$ or $C_v$ used in the general calculation formula.

(b) Laminar flow and excessive flow

When working under non-prime flow conditions, the calculation formula of Newtonian fluid flow through the control valve is derived from the basic formula in GB/T 17213.1-2015, which is applicable to $Re_v$ ,<1000.

The flow coefficient shall be determined by the following formula

$$C = \frac{q_v}{N_1 F_R} \sqrt{\frac{\rho_1 \rho_0}{\Delta p}} \qquad (7)$$

## 3 CONTROL VALVE DATA MODEL

least squares support vector machine[6-8]

Consider the training set of data samples $\{(x_i, y_i)\}$, input data $x_i \in R^n$, output data $y_i \in R$, $i = 1, 2, \ldots, l$. The 10 least squares support vector machine maps the data set to the feature space by nonlinear mapping and transforms the problem into a high-dimensional space fitting. According to the structural risk minimization, the inequality constraint is changed into equality constraint, and the empirical risk is changed from the first power to the second power, and the regression problem is transformed into the constraint optimization problem:

$$\begin{aligned} &\min \frac{1}{2}(\omega \cdot \omega) + \frac{1}{2} C \sum_{i=1}^{l} \xi_i^2 \\ &\text{s.t.} \quad y_i - \omega^T \varphi(x_i) - b = \xi_i \\ &\qquad i = 1, 2, \cdots, l \end{aligned} \qquad (8)$$

Lagrange function is established for equation (8)

$$\mathbf{L} = \frac{1}{2}(\omega \cdot \omega) + \frac{1}{2}C\sum_{i=1}^{l}\xi_i^2 - \sum_{i=1}^{l}\alpha_i[\omega^T\varphi(x_i) + b + \xi_i - y_i] \tag{9}$$

Where, $\alpha_i$ is the Lagrange multiplier..

take derivatives of $\omega, b, \xi_i, \alpha_i$ respectively

$$\begin{aligned}
\frac{\partial L}{\partial \omega} &= 0 \rightarrow \omega = \sum_{i=1}^{l}\alpha_i\varphi(x_i) \\
\frac{\partial L}{\partial b} &= 0 \rightarrow \sum_{i=1}^{l}\alpha_i = 0 \\
\frac{\partial L}{\partial \xi_i} &= 0 \rightarrow \alpha_i = c\xi_i \\
\frac{\partial L}{\partial \alpha_i} &= 0 \rightarrow \omega^T\varphi(x_i) + b + \xi_i - y_i = 0
\end{aligned} \tag{10}$$

And write equation (10) in matrix form

$$\begin{bmatrix} 0 & 1^T \\ 1 & K + C^{-1}I \end{bmatrix}\begin{bmatrix} b \\ \alpha \end{bmatrix} = \begin{bmatrix} 0 \\ Y \end{bmatrix} \tag{11}$$

Where ; $K_{i,j} = (\varphi(\mathbf{x}_i) \cdot \varphi(\mathbf{x}_j)) = k(\mathbf{x}_i, \mathbf{x}_j)$; $k(\bullet, \bullet)$ is the kernel function; $i = 1, 2, \ldots, l$; $1 = (1, \ldots, 1)^T$; $Y = (y_1, y_2, \ldots, y_l)^T$; $\alpha = (\alpha_1, \alpha, \ldots, \alpha_l)^T$;

The lS-SVM fitting function is obtained as follows:

$$f(\mathbf{x}) = \sum_{i=1}^{l}\alpha_i k(\mathbf{x}, \mathbf{x}_i) + b \tag{12}$$

$H = K + C^{-1}I$, you can get

$$\begin{aligned}
\alpha &= H^{-1}\left(Y - 1 \times \frac{1^T H^{-1} Y}{1^T H^{-1} 1}\right) \\
b &= \frac{1^T H^{-1} Y}{1^T H^1 1}
\end{aligned} \tag{13}$$

## 4 CONTROL VALVE DATA MODEL

The mixing model of regulating valve was established

In this paper, an analytical model-based approach is used for intelligent decision making and state monitoring, whose accuracy depends to a certain extent on high-quality process models. Process models generally fall into the following categories: mechanism models, knowledge-based models, and data-based models.[9-13]

Table 1 Comparison of advantages and disadvantages of mechanism model, knowledge model and data-driven model

| method | advantage | disadvantage |
| --- | --- | --- |
| mechanism model | Reflect the internal structure and mechanism of the process, interpolation, extension and portability are good. | For complex systems, modeling is difficult. |
| knowledge model | Only production experience and process knowledge as a priori knowledge, easy to get. | Poor generality, uncertain scheme and countermeasures, difficult to deal with the unknown situation. |
| Data-driven model | Only process data under normal working conditions are needed, and the method is simple. | Poor extensibility, large amount of data required, easy overfitting, and unclear physical meaning of variables. |

Psichogins and Ungar proposed in 1991 a hybrid modeling technique combining mechanism model and data-driven model, which is complementary to each other in terms of both the explicit physical meaning of the model and the high accuracy of the model, as well as good local approximation performance and good global performance.

Assume a nonlinear dynamic process model as follows:

$$\begin{aligned} x(k+1) &= f(x(k), u(k), d(k)) \\ y(k) &= g(x(k)) \end{aligned} \quad (14)$$

Where, x(k) is the state quantity, u(k) is the input quantity, d (k) is the disturbance quantity, ylk is the output quantity, and f(x) and G (x) are all nonlinear functions.In the actual process, f(x) and g (x) are difficult to be directly obtained. The idea of hybrid model is to conduct mechanism modeling for the known mechanism and part of the system, and identify the unknown mechanism part of the system with data-driven modeling method. The mechanism model and data-driven model complement each other.That is to say, on the one hand, the mechanism model provides the prior knowledge of the process for the data-driven model and reduces the requirements on the sample data; on the other hand, the data-driven model compensates the unmodeled characteristics of the mechanism model.

The mixed model can be divided into three types from the connection mode: series mode, parallel mode and mixed mode . In this paper, the mixed model of regulator valve is established by series model.

In some systems, the structure of the mechanism model is determined, but it contains unknown variables or unknown functions, which limits the application of the mechanism model in the actual system. If the measurement data of the system can be used to identify the unknown part of the mechanism model and improve the degree of "whitening" of the system, the accuracy of the model will be greatly improved, and the process control and fault detection and diagnosis will be convenient.

Suppose the nonlinear function f (x) in formula (14) contains the nonlinear parameter P, there is

$$f(x) = k(x, u, p, d) \qquad (15)$$

Among them, the nonlinear parameter P can hardly be obtained from the mechanism model (such as the equilibrium equation)

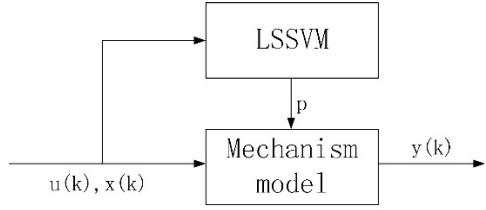

Fig. 3 Series model of regulator valve

$$Q = C_v \varepsilon \frac{f(x)}{\sqrt{1-\beta^4}} \sqrt{\frac{2(p_1 - p_c)}{\rho_1}} \qquad (16)$$

The structure of the series hybrid model with the combination of mechanism and data is given. After the prior knowledge is used to establish the mechanism model of the system, the nonlinear parameter P is identified by the data-based method, and the identified parameter P is substituted into the mechanism model, thus the series mixed model is established. Generally speaking, as long as the structure of the mechanism model is determined, the series hybrid model can be used to model the complex system with strong nonlinear. However, when there are too many nonlinear parameters, the time consumption and algorithm complexity will also increase.

## 4 Simulation Analysis

The DAMADICS actuator fault reference simulation platform is The research object is shown in Figure 3. The DAMADICS Benchmark process control system is used to simulate various types of failures and their characteristics in Table 2 .

Table 1 The fault simulation of benchmark actuator system

|     | fault description | fault direction | fault type |
| --- | --- | --- | --- |
| | Control valve faults | | |
| f1 | Valve clogging | <0, 1> | abrupt |
| f2 | Valve plug or valve seat sedimentation | <0, 1> | slowly developing |
| f3 | Valve plug or valve seat erosion | <0, 1> | slowly developing |
| f4 | Increased of valve or bushing friction | <-1, 1> | slowly developing |
| f5 | External leakage (leaky bushing, covers, terminals) | <0, 1> | slowly developing |
| f6 | Internal leakage (valve tightness) | <0, 1> | slowly developing |
| f7 | Medium evaporation or critical flow | <0, 1> | abrupt |
| | Pneumatic servo-motor faults | | |
| f8 | Twisted servo-motor's piston rod | <0, 1> | abrupt |
| f9 | Servo-motor's housing or terminals tightness | <0, 1> | abrupt |
| f10 | Servo-motor's diaphragm perforation | <0, 1> | abrupt |
| f11 | Servo-motor's spring fault | <0, 1> | abrupt |
| Positioner faults | | | |
| f12 | Electro-pneumatic transducer fault | <-1, 1> | abrupt |
| f13 | Rod displacement sensor fault | <-1, 1> | slowly developing |
| f14 | Pressure sensor fault | <-1, 1> | abrupt |
| f15 | Positioner feedback fault | <0, 1> | abrupt |
| General faults / external faults | | | |
| f16 | Positioner supply pressure drop | <0, 1> | rapidly developing |
| f17 | Unexpected pressure change across the valve | <-1, 1> | rapidly developing |
| f18 | Fully or partly opened bypass valves | <0, 1> | abrupt |
| f19 | Flow rate sensor fault | <-1, 1> | abrupt |

To measure the prediction performance of LS - SVM model, root mean square error was selectedRMSE, MEAN absolute percentage error MAPE and maximum absolute percentage error . For evaluation index, see the definition following

(1) mean-root-square error

$$\mathrm{RMSE} = \sqrt{\frac{1}{n}\sum_{i=1}^{n}(y_i - \hat{y}_i)^2} \quad (18)$$

(2) Mean Absolute percentage error

$$\mathrm{MAPE} = \frac{1}{n}\sum_{i=1}^{n}\frac{|(y_i - \hat{y}_i)|}{y_i} \times 100\% \quad (19)$$

(3) Maximum absolute percentage error

$$\mathrm{Err}_{max} = \max\left(\frac{|y_i - \hat{y}_i|}{y_i} \times 100\%\right) \quad (20)$$
$$i = 1,\cdots,n$$

Where, $n$ is the sample number; $y_i$ is the actual output; $\hat{y}_i$ is the predicted output.

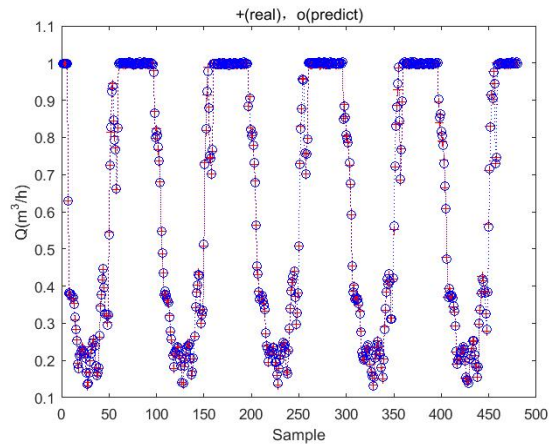

Fig. 4. Simulation results of Feature P1,P2,X

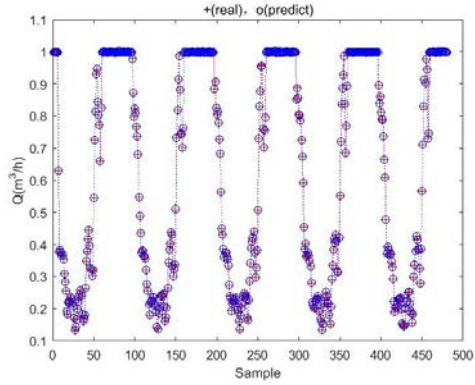

Fig. 5. Simulation results of Feature P1,P2,X,T

Table 2  Results analysis

| Input eigenvector | Flow prediction | | |
|---|---|---|---|
| | RMSE / $m^3 \cdot h^{-1}$ | MAPE / % | $Err_{max}$ / % |
| P1,P2,X | 0.34 | 6.3 | 3.84 |
| P1,P2,X,T | 3.11 | 0.25 | 2.73 |

**Conclusion**

The input feature vector can be seen from FIG. 4 and 5 that the mixed model of regulating valve flow has good accuracy and is related to the feature input.The hybrid modeling method designed in this paper has good precision.